\documentclass[letterpaper, 10pt, conference]{IEEEtran}
\usepackage{soul}
\usepackage{url}
\usepackage{subfigure}
\usepackage{graphicx}
\usepackage{multirow}
\usepackage{array}
\usepackage{booktabs}
\usepackage{amssymb}
\usepackage[usenames]{color}
\usepackage[usenames,dvipsnames,table]{xcolor}
\usepackage{comment}
\usepackage{mathptmx}
\usepackage{latexsym}
\usepackage{amssymb}
\usepackage{wrapfig}
\usepackage{times}
\usepackage{epsfig}
\usepackage{subfigure}
\usepackage{ifpdf}
\usepackage{lipsum}
\usepackage{bbding}

\usepackage{glossaries}
\newacronym{ai}{AI}{Artificial Intelligence}
\newacronym{uav}{UAV}{Unmanned Aerial Vehicles}
\newacronym{uuv}{UUV}{Unmanned Underwater Vehicles}
\newacronym{mavs}{MAVs}{Micro Aerial Vehicles}
\newacronym{mav}{MAV}{Micro Aerial Vehicle}
\newacronym{api}{API}{Application Programming Interface}
\newacronym{cps}{CPS}{Cyber-Physical System}
\newacronym{dos}{DoS}{Denial of Service}
\newacronym{scada}{SCADA}{Supervisory Control and Data Acquisition}
\newacronym{ict}{ICT}{Information and Communications Technology}
\newacronym{lti}{LTI}{Linear Time Invariant}
\newacronym{lqg}{LQG}{Linear Quadratic Gaussian}
\newacronym{lqr}{LQR}{Linear Quadratic Regulator}
\newacronym{ncs}{NCS}{Networked-Control System}
\newacronym{sdn}{SDN}{Software Defined Networking}
\newacronym{gan}{GAN}{Generative Adversarial Network}
\newacronym{qgan}{QGAN}{Quantum Generative Adversarial Network}

\usepackage{hyperref}
\hypersetup{
  colorlinks=true,
  linkcolor=black,
  urlcolor=blue!70!black,
  citecolor=blue!70!black,
}
\usepackage{siunitx}
\usepackage{braket}
\input{Qcircuit}

\newtheorem{thm}{Theorem}

\newtheorem{defn}[thm]{Definition}

\usepackage[normalem]{ulem}
\usepackage{color}

\usepackage{scalerel}
\usepackage{tikz}
\usetikzlibrary{svg.path}
\definecolor{orcidlogocol}{HTML}{A6CE39}
\tikzset{
	orcidlogo/.pic={
		\fill[orcidlogocol] svg{M256,128c0,70.7-57.3,128-128,128C57.3,256,0,198.7,0,128C0,57.3,57.3,0,128,0C198.7,0,256,57.3,256,128z};
		\fill[white] svg{M86.3,186.2H70.9V79.1h15.4v48.4V186.2z}
		svg{M108.9,79.1h41.6c39.6,0,57,28.3,57,53.6c0,27.5-21.5,53.6-56.8,53.6h-41.8V79.1z M124.3,172.4h24.5c34.9,0,42.9-26.5,42.9-39.7c0-21.5-13.7-39.7-43.7-39.7h-23.7V172.4z}
		svg{M88.7,56.8c0,5.5-4.5,10.1-10.1,10.1c-5.6,0-10.1-4.6-10.1-10.1c0-5.6,4.5-10.1,10.1-10.1C84.2,46.7,88.7,51.3,88.7,56.8z};
	}
}
\newcommand\orcidicon[1]{\href{https://orcid.org/#1}{\mbox{\scalerel*{
				\begin{tikzpicture}[yscale=-1,transform shape]
				\pic{orcidlogo};
				\end{tikzpicture}
			}{|}}}}

\begin{document}

\title{Faking and Discriminating the Navigation Data of a Micro Aerial Vehicle Using Quantum Generative Adversarial Networks}

\author{\IEEEauthorblockN{Michel Barbeau}
\IEEEauthorblockA{Carleton University, Canada}
\orcidicon{0000-0003-3531-4926}{0000-0003-3531-4926}
\and
\IEEEauthorblockN{Joaquin Garcia-Alfaro}
\IEEEauthorblockA{Telecom SudParis, France}
\orcidicon{0000-0002-7453-4393}{0000-0002-7453-4393}
}

\thanks{M. Barbeau, Carleton University, School of Computer Science,
Canada. E-mail: barbeau@scs.carleton.ca}

\thanks{J. Garcia-Alfaro, Institut Polytechnique de Paris,
CNRS UMR 5157 SAMOVAR, T\'el\'ecom SudParis, France. E-mail:
jgalfaro@ieee.org}

\thanks{Manuscript received June, 2019; revised xxxx.}

\maketitle

\thispagestyle{plain}
\pagestyle{plain}

\begin{abstract}
We show that the \gls*{qgan} paradigm can be employed by an adversary to learn  generating data that deceives the monitoring of a \gls*{cps} and to perpetrate a covert attack. 
As a test case, the ideas are elaborated considering the navigation data of a \gls*{mav}.
A concrete \gls*{qgan} design is proposed to generate 
fake \gls*{mav} navigation data. Initially, the adversary is entirely ignorant about the dynamics of the \gls*{cps}, the strength of the approach from the point of view of the bad guy. A design is also proposed to discriminate between genuine and fake \gls*{mav} navigation data. The designs combine classical optimization, qubit quantum computing and photonic quantum computing. Using the PennyLane software simulation, they are evaluated over 
a classical computing platform. We assess the learning time and accuracy of the navigation data generator and discriminator versus space complexity, i.e., the amount of quantum memory needed to solve the problem. 
~~
\end{abstract}

\section{Introduction}

\noindent \gls*{cps}s comprise physical processes monitored and
controlled through embedded computing and networked resources. 
Signals to actuators and
feedback from sensors are exchanged with controllers using, e.g.,
wireless communication. The advantages of such architectures include
flexibility and relatively low deployment costs. Nevertheless, the perpetration of cyber-physical attacks must be addressed. The 
problem is 
particularly challenging when the \gls*{cps} consists of disruptive 
technologies such as \gls*{mav}s, \gls*{uav}, \gls*{uuv} and \gls*{mav} 
swarming.

Today's cybersecurity solutions, from in-depth defense techniques
(e.g., firewalls) to intrusion detection and cryptographic techniques,
aim to prevent system breaches from happening. However, several
stories of attacks and disruption of \gls*{cps} exist (e.g., from the
\href{http://j.mp/2jaM6uM}{Stuxnet worm} incident affecting a Iran's
\emph{atomic program}~\cite{stuxnet} to recent incidents 
in Saudi Arabia affecting \href{http://bit.ly/2LMqR3H}{Houthi drones}~\cite{houthiDrone}). \gls*{cps} protection solutions 
must manage and take control over adversarial actions.
Protection must be built taking on the adversary mindset,
predicting its intentions and adequately mitigating the effects of
its actions. 

In this paper, we explore the use of the  \gls*{qgan} paradigm to address cyber-physical security issues in the domain of \gls*{mav}s.
A concrete \gls*{qgan} design is proposed to generate 
fake \gls*{mav} navigation data. Initially, the adversary is entirely ignorant about the dynamics of the \gls*{cps}. From the point of view of the adversary, it is the strength of the approach. A design is also proposed to discriminate between real and fake \gls*{mav} navigation data. The designs combine classical optimization, qubit quantum computing and photonic quantum computing. 
We build upon the PennyLane quantum machine learning software platform~\cite{Bergholm2018}. In particular, we reuse and adapt ideas 
from the variational
classifier~\cite{Schuld2019Q3} and \gls*{qgan}~\cite{Schuld2019Q4}
examples. 

We evaluate our approach using the simulation capabilities of PennyLane. We measure the learning time and accuracy of the navigation data generator and discriminator with respect to the space complexity, i.e., the amount of quantum memory used to solve the problem. 
At the outset, we acknowledge that the exponentially growing time complexity in the number of qubits of our solution is a barrier to its application on a large scale.
In particular, when the calculations are all done in simulation over a classical computing platform.
Nevertheless, we show the feasibility of the approach on a small scale and identify hurdles that are likely to be overcome by the upcoming evolution of quantum machine learning.

\medskip

Section~\ref{sec:problem} elaborates further on our problem domain
and related work. Section~\ref{sec:approach} presents our solution.
Section~\ref{sec:performance} provides experimental work.
Section~\ref{sec:conclusion} concludes the paper\footnote{Accepted for publication in IEEE GLOBECOM 2019 Workshop on Quantum Communications and Information Technology 2019 (fifth QCIT workshop of the \textit{Emerging Technical Committee on Quantum Communications and Information Technology}, QCIT-ETC, cf. \url{http://qcit.committees.comsoc.org/qcit19-workshop/}).}.

\section{Problem Domain}
\label{sec:problem}

The problem domain encompasses \gls*{cps} controllers, playing the role of defenders, and adversaries. We conceptualize the situation in
terms of activities consisting of gathering and hiding knowledge about both defensive and adversarial strategies. We
envision the use of new learning theories, in which defenders and
adversaries conceal their actions to avoid being profiled for the purpose
of thwarting their \emph{cyber-physical
battle weapons}. 
Defenders equipped with \gls*{ai} tools, such as machine
learning, can be identify adversarial actions trying to collect as much knowledge as possible about their targets. The defense starts
by learning the weaknesses of the adversaries and offensively
mislead their intentions, thwarting their actions in the
end. Once the defender knows the adversary, e.g., the
behavior performed to identify and disrupt the services,
the defender starts offering assets sacrificed to coax the
adversary and to manage a potential security breach.

We are particularly interested in a type of \gls*{cps} which function is air space 
surveillance and coastal water monitoring. The application domain of interest
includes \gls{mavs} and
related technologies such as \gls*{uav}, \gls*{uuv},
formations of \gls*{mavs} and collaborating \gls*{mavs}.
We focus on
scenarios where an adversary targets the components of the
\gls*{cps} and perpetrates covert cyber-physical attacks~\cite{Teixeira2015,smith2015covert}. 
The adversary is to operate a stealthy disruption of 
services.
The purpose is disrupting the
navigation data of the \gls{mavs} and deceive the defender. The
role of the defender is to recognize the activities performed by the
adversary, i.e., identify the intentions of the adversary and
correct the adversarial actions.

\subsection{Covert Attack and Feedback Truthfulness}

A covert attack is an aggression on the state of a \gls*{cps} where
the adversary attempts to be invisible~\cite{Teixeira2015}. It is
assumed that the adversary knows or can learn the dynamics of 
the \gls*{cps}. While the attack is being carried out, the perpetrator compensates the impact of the attack over the system by providing fake information to the system operators (e.g., by concealing the
effect of the spoofed inputs). Hence, from the point of view of an
observer, responsible for detecting the attack, the execution of
the \gls*{cps} looks normal. 
\begin{figure}[!b]
\subfigure[]{
\includegraphics[width=.45\columnwidth]{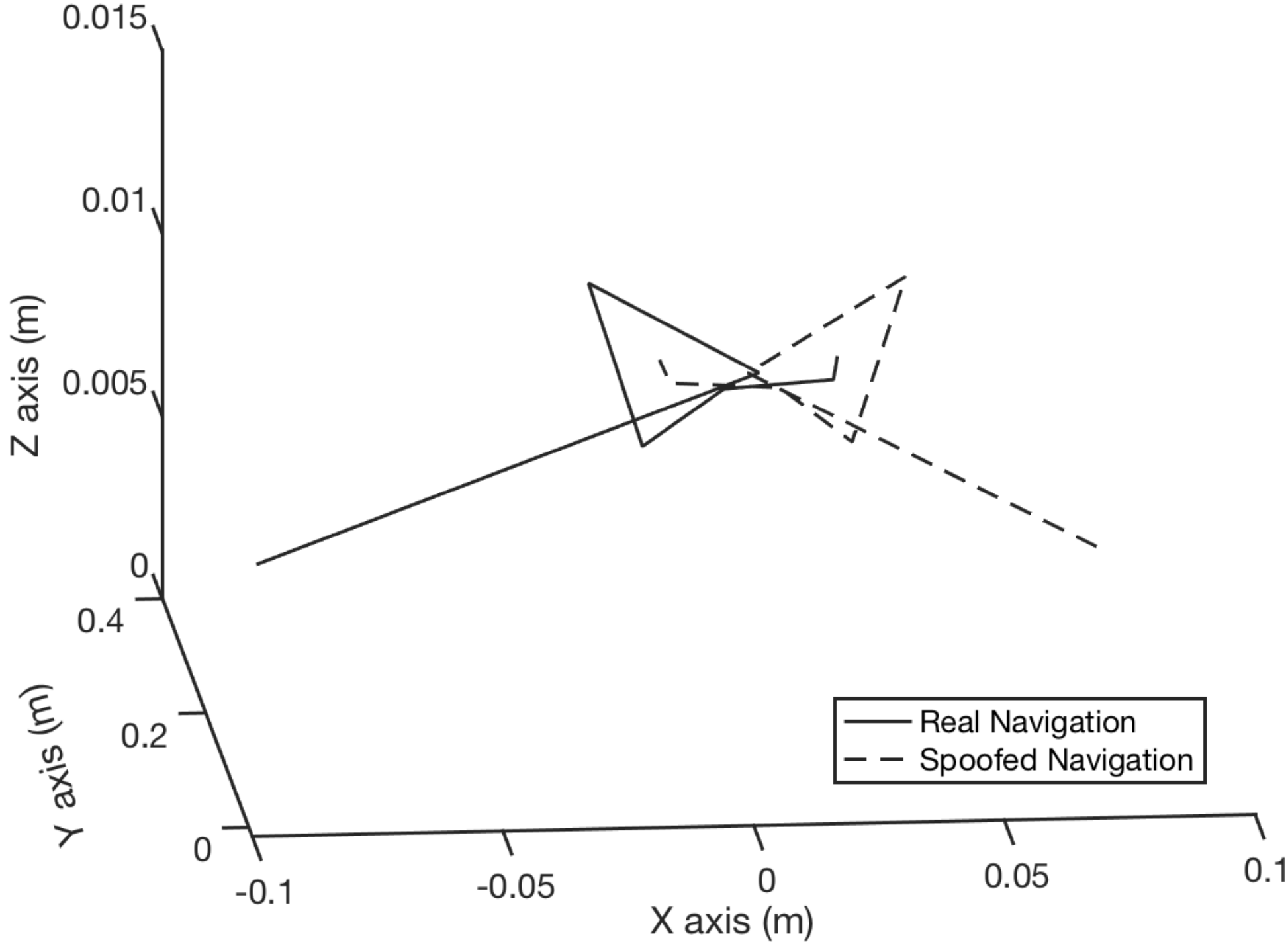}
}
\subfigure[]{
\includegraphics[width=.45\columnwidth]{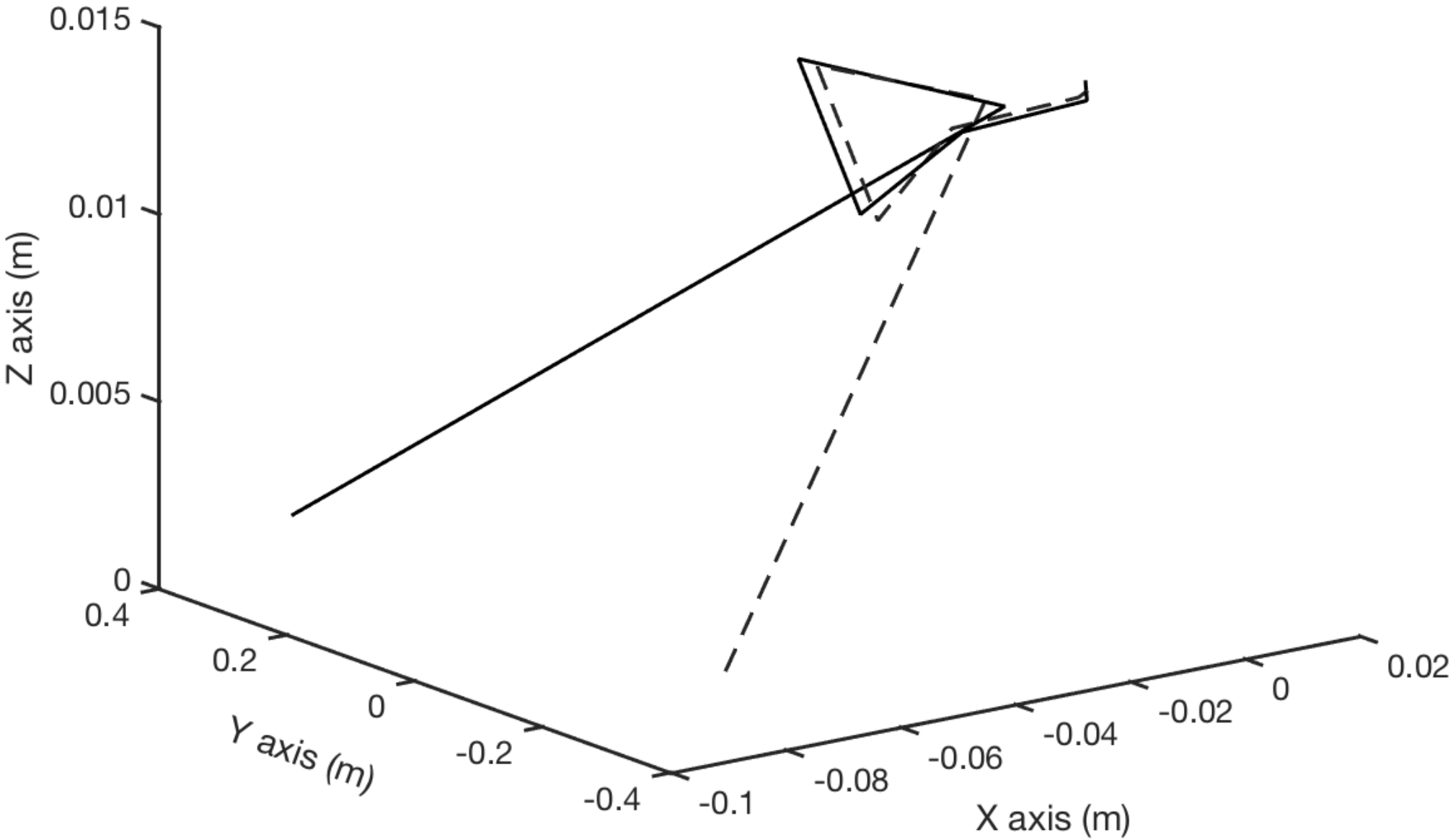}
}
\subfigure[]{
\includegraphics[width=.45\columnwidth]{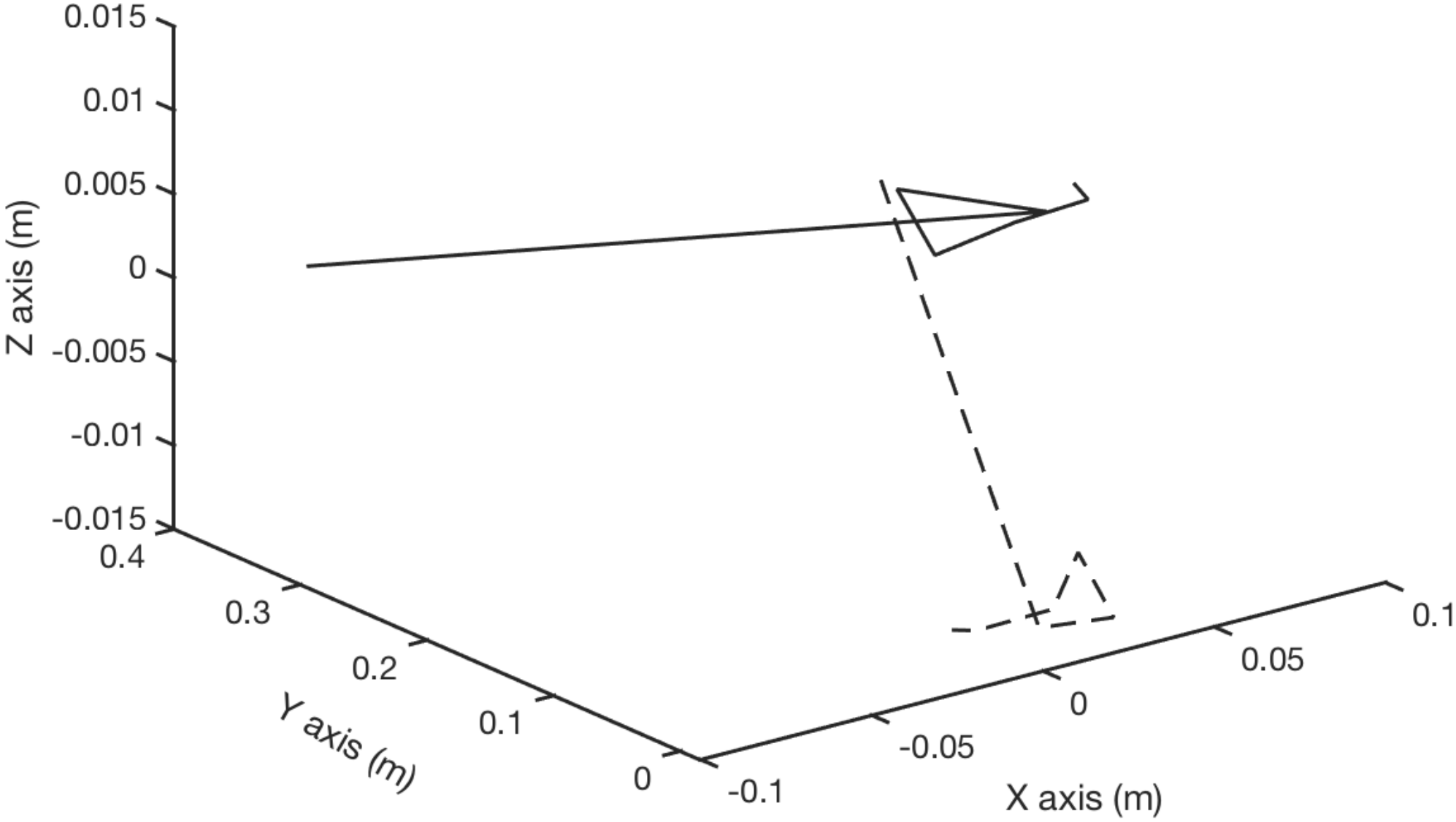}
}
\subfigure[]{
\includegraphics[width=.45\columnwidth]{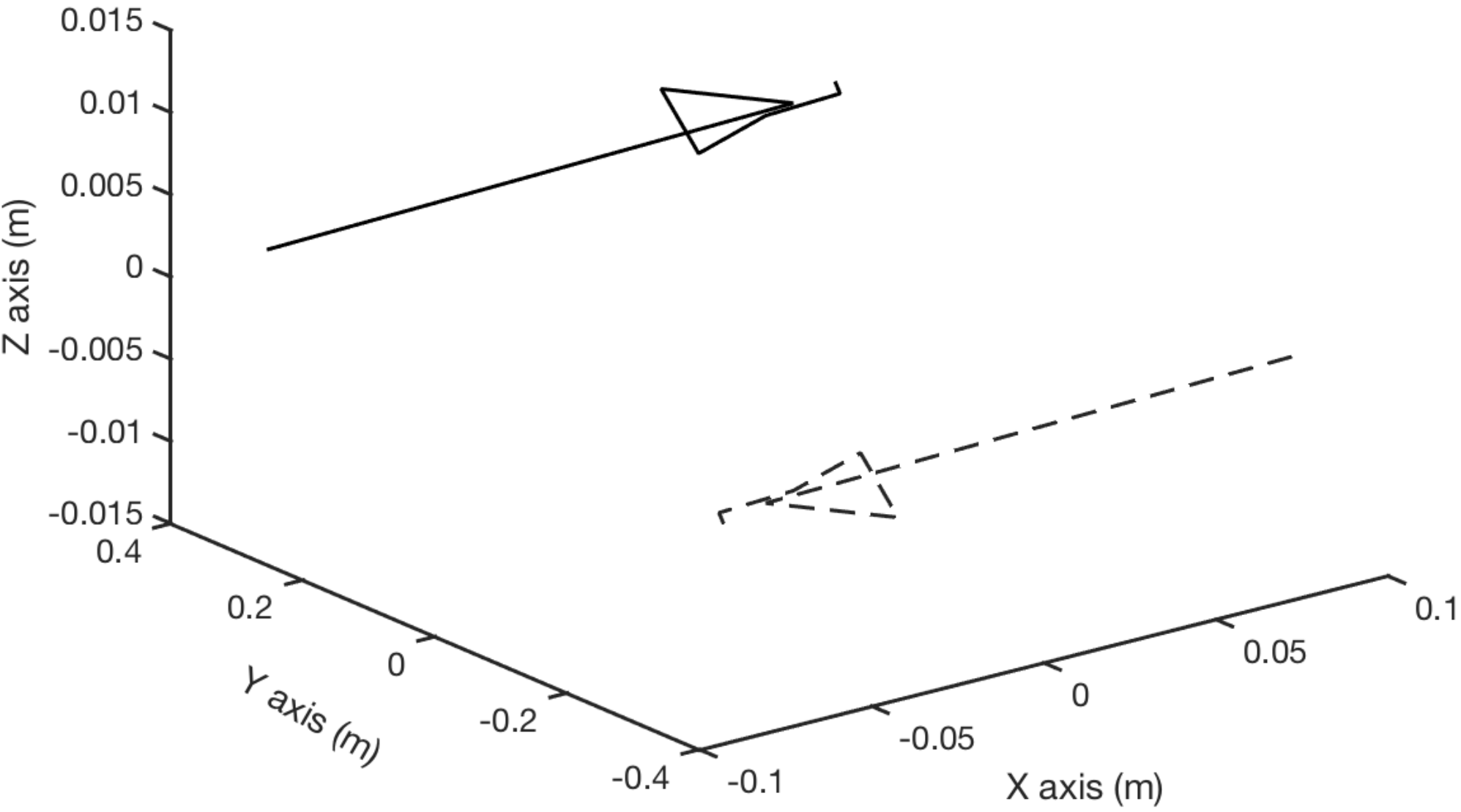}
}
\caption{\gls*{mav} navigation data trace disruptions. The adversary 
perpetrates GPS-like attacks~\cite{barbeau2019geocaching,javaid2017analysis},
  e.g., to swap the navigation coordinates. Solid lines represent genuine navigation data. Dashed lines represent disrupted navigation data.
  (a)~Swapping of the $x$ coordinate. (b)~Swapping of the $y$ coordinate. (c)~Swapping of the $x,z$ coordinates. (d)~Swapping of the $x,y,z$ coordinates. \label{fig:fig1}}
\end{figure}
Assume the scenario shown in Figure~\ref{fig:fig1}. It depicts the
disruption of the navigation data of a series of \gls*{mavs}. The
manipulation is conducted by a remote adversary via, e.g., GPS jamming
and spoofing attacks~\cite{barbeau2019geocaching,javaid2017analysis}.
The goal of the adversary is to conduct navigation data modifications
(e.g., swapping the $x,y$ coordinates of the navigation traces) and
hide the disruption to the defender, with additional cyber-physical
covert attacks~\cite{Teixeira2015, smith2015covert}.

\begin{figure}[!t]
\centering
\subfigure[]{
\includegraphics[width=0.5\columnwidth]{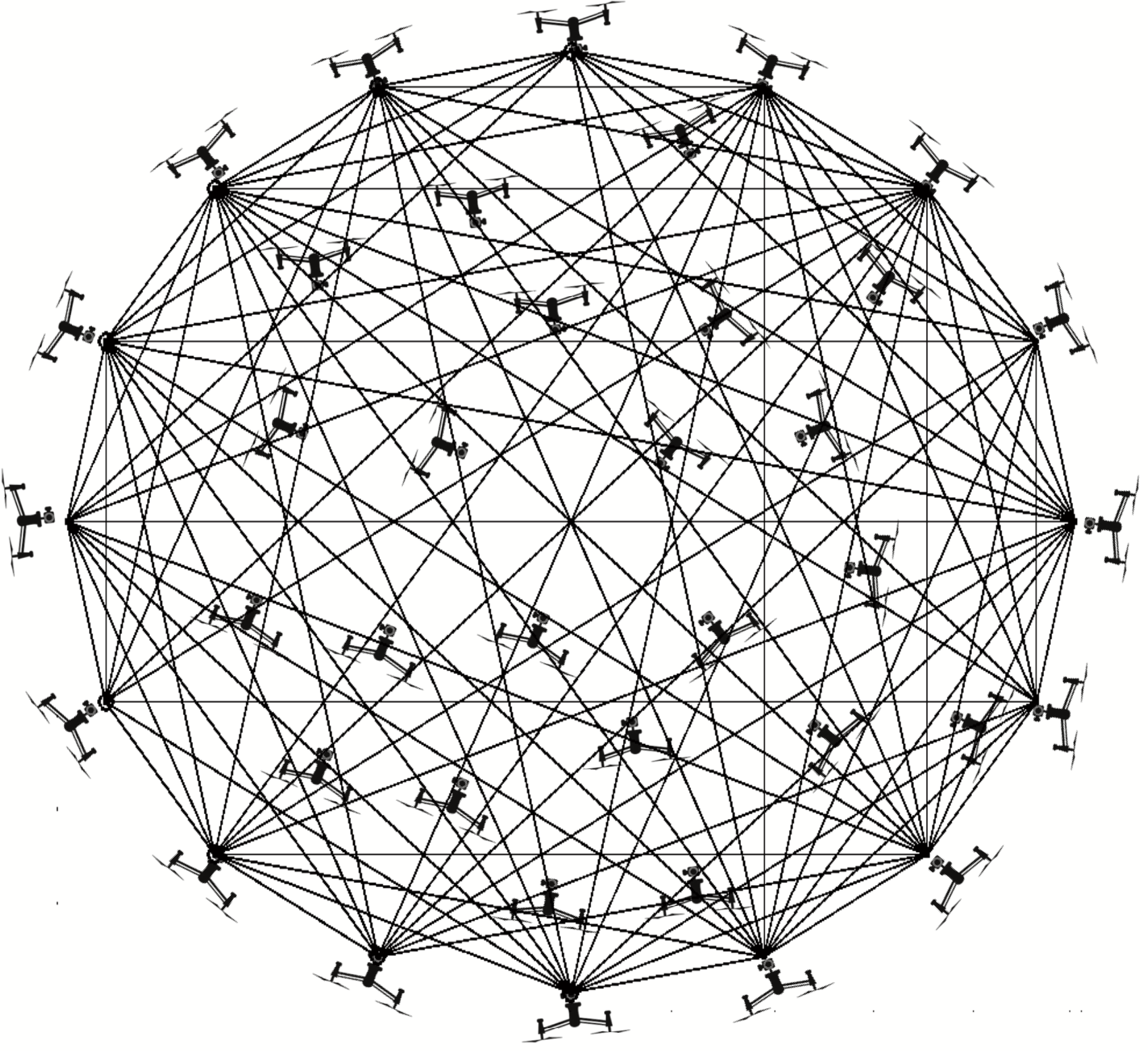}
}
\subfigure[]{
\includegraphics[width=0.5\columnwidth]{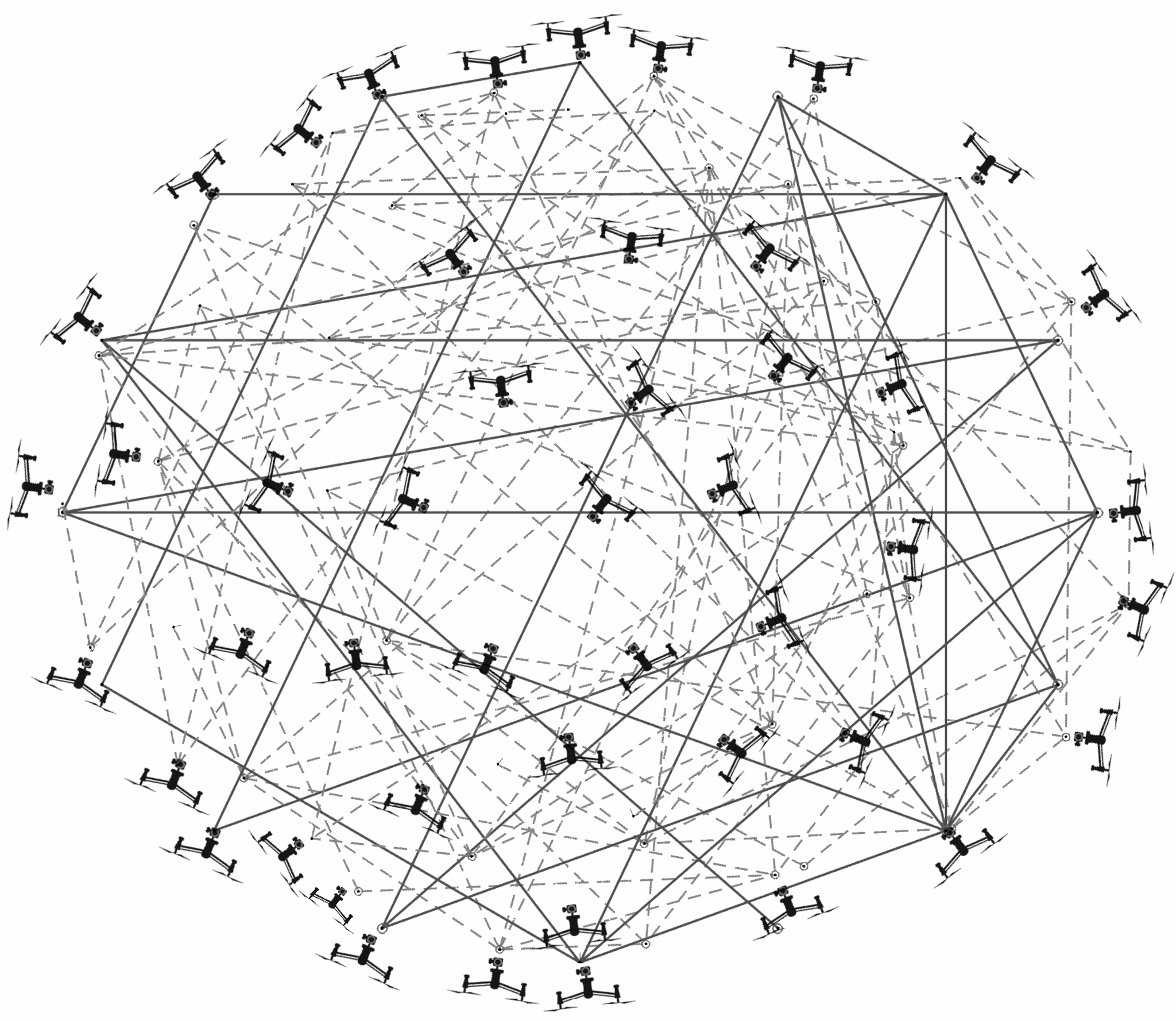}
}
\caption{(a)~ \gls*{mav} genuine navigation data. (b)~Extract of some \gls*{mav}
  navigation data disrupted by GPS attacks (solid lines 
  represent genuine navigation data; dashed lines represent spoofed navigation data).\label{fig:fig2}}
\end{figure}
Assume now the merge of multiple navigation traces of a single \gls*{mav} over a given period of time, as depicted in
Figure~\ref{fig:fig2}. The expected navigation data seen by the \gls*{cps}
controller is depicted in Figure~\ref{fig:fig2}(a). The spoofed
navigation data, due to the adversarial attacks, is depicted  by the 
dashed lines in Figure~\ref{fig:fig2}(b). The covert attack conducted 
by the adversary conceals the alteration of some of the navigation 
paths. The defender conducts a learning process to guess the adversarial
intentions. The defender also prioritizes assets that can get
sacrificed as collateral damages (e.g., to offer some tactical
victories to the adversary with the aim of reducing the adversarial
power in the long term), e.g., by using some game-theoretic ideas. 
The process allows the defender to get trusted by the adversary, i.e., 
to make the adversary confident about the success of some 
perpetrated actions. Practically speaking, the collateral damages 
allow the defender to reinforce the defensive learning processes, 
with the aim of handling and correcting the affected system 
represented by Figure~\ref{fig:fig2}(b), to the original plans before 
the execution of the adversarial actions (i.e., Figure~\ref{fig:fig2}(a)). Both graphs (the one seen by the defender and the one seen by 
the adversary) evolve dynamically over time, and converge 
eventually. Successful victories of the learning process
conducted by the defender increases the converge likelihood 
of the two graphs.

\subsection{Related Work}

Related work include the use of machine learning for cyber-physical 
protection and extended machine learning functions relying on 
quantum techniques. Details  follow.

\subsubsection{\gls*{cps} Protection Using Machine 
Learning} 

The domain of \gls*{ai}, by means of the 
subfields of search and machine learning, provides a large set of 
techniques relevant to the resilience of a \gls*{cps}.  Supervised, unsupervised and
reinforcement
are the three main machine
learning paradigms. In supervised machine learning, there are old and new
data points. Old data points are labelled, representing
classes of data points. Comparing their similarity with old
ones, supervised machine leaning assigns labels to
new data points. With unsupervised machine learning, the data points
are unlabelled. Learning  consists of extracting information from
data. Data points are grouped together into classes according to
similarity. Human experts label the classes. 

In contrast, reinforcement learning rewards or penalizes the learner
following the validity of inferred classifications. Learning is  from the successes 
and mistakes. Supervised and reinforcement machine learning is
used for system identification and model fitting. 
Different alternative learning
methods exist, based on different considerations on the type of model
(e.g., rule-based, support-vector machines, deep learning models) and
its properties (e.g., explainable models/decisions, efficiency). The
perpetration of control-theoretic attacks 
\cite{Teixeira2015,smith2015covert} may require a
system identification phase performed by the adversary. Kernels
methods~\cite{shawe2004kernel}, a kind of machine learning, can be used for system identification~\cite{PILLONETTO2014657,PILLONETTO201081}.

\subsubsection{The Quantum Advantage} 

The time complexity of quantum search
techniques are data size independent. Along the same line, 
quantum machine learning, i.e.,
the use of quantum computing for machine learning, has great potential
because the time complexity of classification is independent of the number
of data points. Schuld and Killoran investigated the use of kernel
methods~\cite{shawe2004kernel}, that can be used for system
identification, for quantum machine
learning~\cite{schuld2018supervised,Schuld2019}. 
Encoding of classical
data into a quantum state is needed. A similar approach has been
proposed by Havl{\'\i}{\v c}ek et al.~\cite{havlicek2019}. 

Schuld and
Petruccione~\cite{schuld2018supervised} discussed in details the application
of quantum machine learning classical data generation and quantum data
processing. A translation procedure is required to map the classical
data, i.e., the data points, to quantum data, enabling quantum data
processing, i.e., quantum classification. However, there is a cost
associated with translating classical data into the quantum form,
which is comparable to the cost of classical machine learning
classification. This is right now the main barrier. The approach that will
result in real gains is quantum data generation and quantum data
processing, there will be no need to translate from classical to quantum
data. Quantum data generation requires quantum sensing.

\section{Faking and Discriminating Navigation Data}
\label{sec:approach}
Using the \gls*{gan} framework, we validate that a covert attack
can be perpetrated using adversarial learning. A \gls*{gan} consists
of two main entities: a discriminator and a
generator~\cite{NIPS2014_5423}. The discriminator is the defender's
tool. The generator is the adversary tool. There are genuine (real) data and
generated (fake) data. The generator aims at generating data to
deceive the discriminator. The discriminator is trained with genuine and
generated data. The training process aims to a discriminator able to 
label genuine or generated data correctly, with high probability 
of correctness. The
adversary wins the game when this probability is at least 50\%. 
To this end,
the generator is trained, assuming it can challenge the discriminator 
with data and access to the verdict.
Training is an iterative process. Training iterates until the production 
of fake data is accepted by the discriminator with high probability.

In a \gls*{qgan}~\cite{Dallaire2018,Lloyd2018} the data can be
quantum. Using a \href{https://www.parrot.com/global/support/products/mambo-fpv}{Parrot Mambo} \gls*{mav}, we generate genuine
navigation data. The navigation is classical and in continuous domains. 
Using probability amplitude encoding, the genuine (classical) data 
is mapped to quantum data and used to train a discriminator, defined as
a qubit-quantum circuit. Using a photonic-quantum circuit, we
validate that the adversary can learn to generate fake data
resembling genuine data, assuming access to nothing else but the verdict
of the discriminator.

\subsection{Discriminator Design}
\label{sec:discriminator}

We build upon the PennyLane~\cite{Bergholm2018} variational
classifier~\cite{Schuld2019Q3} and \gls*{qgan}~\cite{Schuld2019Q4}
examples.
\begin{figure}[h]
\begin{center}
\[
\Qcircuit @C=2em @R=2em {
& & \mbox{Elementary circuit $\mathcal{E(\omega)}$} \\
&\lstick{\ket{\psi_0}} & \gate{Rot(\omega_{0,0},\omega_{0,1},\omega_{0,2})} & \ctrl{1} & \qw      & \targ     & \qw \\
&\lstick{\ket{\psi_1}} & \gate{Rot(\omega_{1,0},\omega_{1,1},\omega_{1,2})} & \targ    & \ctrl{1} & \qw       & \qw \\
&\lstick{\ket{\psi_2}} & \gate{Rot(\omega_{2,0},\omega_{2,1},\omega_{2,2})} & \qw      & \targ    & \ctrl{-2} & \qw
\gategroup{2}{3}{4}{6}{1.7em}{--}
}
\]
\caption{Three-qubit elementary circuit layer.}
\label{fig:elementarycircuit}
\end{center}
\end{figure}
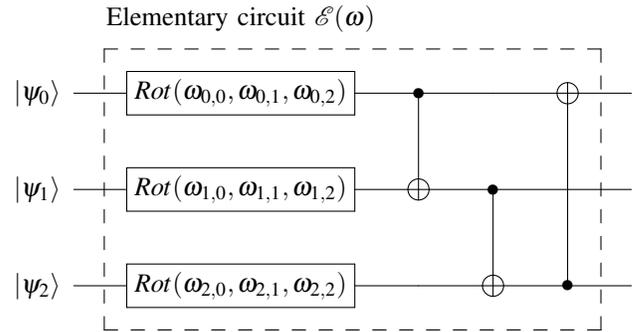
The elementary circuit design $\mathcal{E(\omega)}$ of Farhi and
Neven~\cite{Farhi2018} is used, pictured in
Figure~\ref{fig:elementarycircuit}. Every elementary circuit processes $n$ qubits.
In Figure~\ref{fig:elementarycircuit}, $n$ is three. 
The circuit formal parameter $\omega$ is a $n$ by three matrix of rotation angles.
For $i=0,1,\ldots,n-1$, the gate
$Rot(\omega_{i,0},\omega_{i,1},\omega_{i,2})$ applies the $x$, $y$ and
$z$-axis rotations $\omega_{i,0}$, $\omega_{i,1}$ $\omega_{i,2}$ to qubit $\ket{\psi_i}$. The
three rotations can take a qubit from any state to any state. For entanglement
purposes, qubit $i$ is connected to qubit $i + 1$ modulo $n$ using a
CNOT gate.
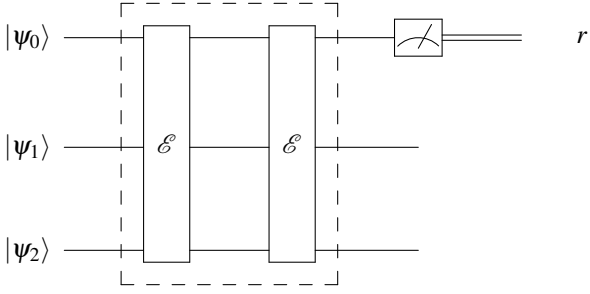
\begin{figure}[h]
\begin{center}
\[
\Qcircuit @C=3em @R=3em {
& & \mbox{Discriminator circuit $\mathcal{D(\omega)}$} & & \\
&\lstick{\ket{\psi_0}} & \multigate{2}{\mathcal{E}}  & \multigate{2}{\mathcal{E}} & \meter & \cw & \lstick{r}\\
&\lstick{\ket{\psi_1}} & \ghost{\mathcal{E}}  & \ghost{\mathcal{E}} & \qw  \\
&\lstick{\ket{\psi_2}} & \ghost{\mathcal{E}} & \ghost{\mathcal{E}} & \qw
\gategroup{2}{3}{4}{4}{1.7em}{--}
}
\]
\caption{Discriminator circuit made of two layered elementary circuits.}
\label{fig:discriminatorcircuit}
\end{center}
\end{figure}
The discriminator circuit $\mathcal{D(\omega)}$ uses $m$ layers of elementary circuits $\mathcal{E}$. 
In Figure~\ref{fig:discriminatorcircuit}, $m$ is two. Layer $0$ accepts
the input. Layer $i$ quantum outputs are connected to layer $i+1$ quantum
inputs.
In this case, the circuit formal parameter $\omega$ is a $m$ by $n$ by three matrix 
of rotation angles.
Layer $i$ is actualized with sub-matrix $\omega_i$.


We use probability amplitude encoding, because we can represent in a given number 
of qubits an exponential number of data points. Probability amplitude encoding 
requires normalized data.
Let $x_0,\ldots,x_{n-1}$ be the data values, their normal form is
$$
v_0=x_0/\mu,\ldots,v_{n-1}=x_{n-1}/\mu
$$
where
$$
\mu = \sqrt{x_0^2 + \ldots +x_{n-1}^2}.
$$

With probability amplitude encoding, up to $2^{n}$ single scalar
values can be represented in probability amplitudes in the input
circuit quantum state. The input quantum state with probability amplitude encoded
data has the following format:
\begin{equation}\label{eq:amplitudeencoding}
\ket{\psi}
=
\sum_{i=0}^{n-1} v_i \ket{i}
\end{equation}

In Figure~\ref{fig:discriminatorcircuit}, layer $m-1$ produces the
output expectation $r$ on line $0$. The output $r$ ranges in the
continuous interval +1 down to -1, respectively corresponding to
qubits $\vert 0 \rangle$ and $\vert 1 \rangle$. Intermediate values
represent superpositions of qubits $\vert 0 \rangle$ and $\vert 1
\rangle$. The output is interpreted as follows. When it is $+1$, the
data is accepted as true. When it is $-1$, the data is rejected and
considered fake. The output $r$ is converted to a probability value,
in the interval $[0,1]$, using the following conversion:
\begin{equation}\label{eq:prob_true}
p = \frac{r+1}{2}.
\end{equation}
When genuine data is submitted on the inputs ($\ket{\psi}$) of the
discriminator, the value $p$ in Eq.~\eqref{eq:prob_true} expresses the
{\em probability of real true} $p_R$. When fake data submitted, the
value $p$ corresponds to the {\em probability of fake true} $p_F$.

We aim to a discriminator that maximizes the probability $p_R$ of
accepting true data while minimizing the probability $p_F$ of
accepting fake data. An optimizer finds a rotation angle
matrix $\omega$ such that the output
of the circuit is approaching $+1$, which corresponds to qubit $\vert
0 \rangle$. Using a gradient descent technique, the optimizer
iterates with genuine data sets and fake data sets.
Gradient descent means that the optimizer tries to minimize the cost represented 
by the difference $p_F-p_R$.
\begin{defn}[Discriminator optimization problem]
Given the quantum input state $\phi$, 
probability amplitude encoding fake navigation data, 
and quantum input state $\psi$, 
probability amplitude encoding genuine navigation data,
training the discriminator $\mathcal{D}(\omega)$ is the optimization problem 
that consists of finding the matrix $\omega$ ($m\times n\times 3$) 
that gives the smallest difference $p_F-p_R$.
\end{defn}

\subsection{Generator Design}
\label{Generator}
The aim of the generator is to produce fake data that is accepted as
true by the discriminator, i.e., the probability of fake true $p_F$ is
as close to one as possible. When training the generator, it is
assumed that the adversary can submit its fake data to the
discriminator and access to the verdict. For the generator,
we investigated the three following designs: (1) a first generator using a
\gls*{mav} model, (2) a qubit-quantum circuit, and (3) 
a photonic quantum circuit. 

\subsubsection{MAV model design}

A detailed model of the \gls*{mav} is built and evolved. For example,
such a model does exist for the \gls*{mav} we are using for our
experiments~\cite{MathWorks2019}. The continuous domain navigation data
generated by the \gls*{mav} model is amplitude-encoded and submitted
to the discriminator. According to the output of the discriminator,
the \gls*{mav} model is fine tuned until a high probability of fake
data acceptance is reached. The challenge with this approach is that
the adversary needs a detailed understanding of the dynamics of the
\gls*{mav}. We aim at a method that does require no knowledge on the
part of the adversary about the \gls*{mav} dynamics. In other words, 
we aim at an automated learning process. Two alternative designs for 
such a purpose follow.

\subsubsection{Qubit-quantum Circuit}

\begin{figure}[h]
\begin{center}
\[
\Qcircuit @C=1.5em @R=3em {
& & \mbox{Generator circuit} & & & & \mbox{Discriminator circuit} & & \\
&\lstick{\ket{0}} & \multigate{2}{\mathcal{E}} & \multigate{2}{\mathcal{E}} & \qw & \multigate{2}{\mathcal{E}}  & \multigate{2}{\mathcal{E}} & \qw & \meter & \cw & \lstick{r}\\
&\lstick{\ket{0}} & \ghost{\mathcal{E}}  & \ghost{\mathcal{E}} & \qw & \ghost{\mathcal{E}} & \ghost{\mathcal{E}} & \qw  \\
&\lstick{\ket{0}} & \ghost{\mathcal{E}}  & \ghost{\mathcal{E}} & \qw & \ghost{\mathcal{E}} & \ghost{\mathcal{E}} & \qw
\gategroup{2}{3}{4}{4}{1.7em}{--}
\gategroup{2}{6}{4}{7}{1.7em}{--}
}
\]
\caption{Generator qubit circuit feeding the discriminator circuit.}
\label{fig:generatorqubitcircuit}
\end{center}
\end{figure}
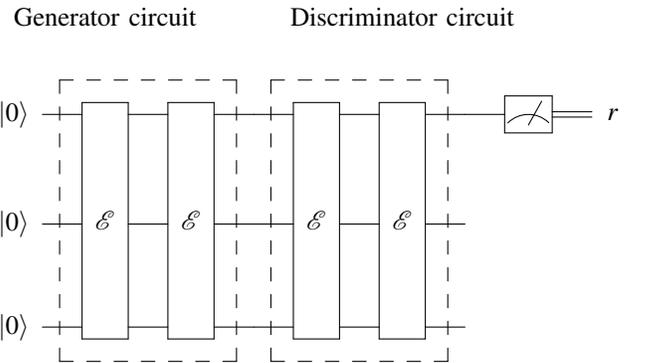
The fake data can be generated with a qubit-quantum circuit, with an
architecture as the one pictured in Figure~\ref{fig:generatorqubitcircuit}. 
The generator circuit is similar to the discriminator circuit, pictured in
Figure~\ref{fig:discriminatorcircuit}. For the generator circuit, the
inputs are all at $\ket{0}$. The optimization is done on the rotation
angles, using the verdict of the discriminator $r$. The learning
process is automatic. The generator outputs the navigation data with
entropy. The outputs of the generator are directly connected to the
inputs of the discriminator. The generated navigation data is encoded
in the probability amplitudes of the quantum state produced by the
generator. Although it works, the navigation data must be transformed 
to a quantum format. Hence, the data is unusable for practically 
perpetrating the attack. Indeed,
qubit-circuit outputs, obtained through measurements, collapse to
zeros and ones. To be usable in an attack scenario, the data needs to
be transformed from classical continuous domains. An alternative 
design for such a purpose follows next.

\subsubsection{Photonic quantum circuit}

The generator combines photonic quantum computing~\cite{Killoran2019} and qubit-quantum computing.
\begin{figure}[h]
\begin{center}
\[
\Qcircuit @C=2.5em @R=2.5em {
& & \mbox{Single-qumode photonic circuit} \\
& \lstick{\ket{0}} & \gate{D(\alpha)} & \gate{R(\phi)} &  \meter & \cw
\gategroup{2}{1}{2}{5}{1.7em}{--}
}
\]
\caption{Photonic-cricuit model.}
\label{fig:photonicdevice}
\end{center}
\end{figure}
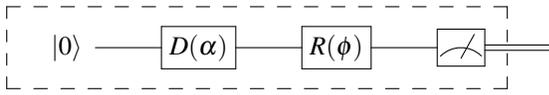
Photonic devices are trained to generate photon numbers corresponding to navigation data accepted
by the discriminator.
A photonic quantum circuit is shown in Figure~\ref{fig:photonicdevice}.
It has a single line, called a {\em qumode}.
The input of the circuit, $\ket{0}$, is the zero energy level. There 
are two Gaussian devices. There is a displacement gate
$D$, with parameter $\alpha$, and a rotation gate $R$, with parameter $\phi$.
They change the circuit energy level and expected numbers of output photons.
The measurement gate determines the average number of photons at the output of the circuit.

We use photonic devices to generate fake navigation data.
The output is amplitude-encoded and submitted to the discriminator.
The photonic-quantum circuit is optimized on the parameters $\alpha$ and $\phi$ such
that the probability of acceptance of the fake data by the discriminator is high.
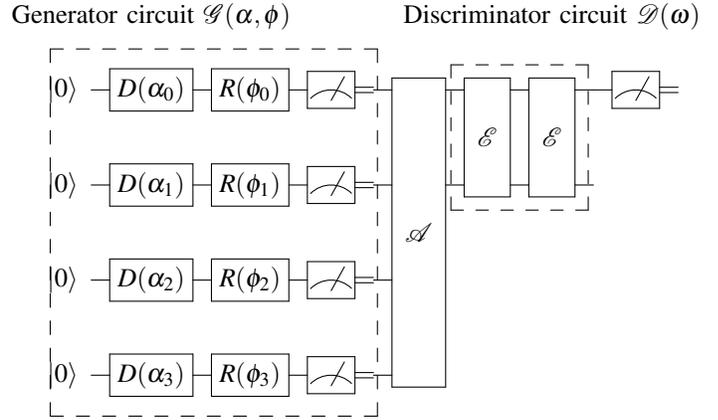
\begin{figure}[h]
\begin{center}
\[
\Qcircuit @C=0.7em @R=2em {
& & \mbox{Generator circuit $\mathcal{G}(\alpha,\phi)$} & & & & & & \mbox{Discriminator circuit $\mathcal{D}(\omega)$} \\
& \lstick{\ket{0}} & \gate{D(\alpha_{0})} & \gate{R(\phi_{0})} &  \meter & \cw & \multigate{3}{\mathcal{A}} & \multigate{1}{\mathcal{E}}  & \multigate{1}{\mathcal{E}} & \qw & \meter & \cw \\
& \lstick{\ket{0}} & \gate{D(\alpha_{1})} & \gate{R(\phi_{1})} &  \meter & \cw & \ghost{\mathcal{A}} & \ghost{\mathcal{E}} & \ghost{\mathcal{E}} & \qw\\
& \lstick{\ket{0}} & \gate{D(\alpha_{2})} & \gate{R(\phi_{2})} &  \meter & \cw & \ghost{\mathcal{A}}\\
& \lstick{\ket{0}} & \gate{D(\alpha_{3})} & \gate{R(\phi_{3})} &  \meter & \cw & \ghost{\mathcal{A}}
\gategroup{2}{1}{5}{5}{1.7em}{--}
\gategroup{2}{8}{3}{9}{1.em}{--}
}
\]
\caption{Generator qubit circuit feeding the discriminator circuit ($n$ is two) .}
\label{fig:generatorqumodetcircuit}
\end{center}
\end{figure}

The architecture pictured in Figure~\ref{fig:generatorqumodetcircuit}
shows a generator feeding a discriminator circuit through a probability amplitude
encoder $\mathcal{A}$, including normalization. The \gls*{mav} navigation 
data set is amplitude encoded according to Eq.~\eqref{eq:amplitudeencoding}. 
Since $n$ qubits can amplitude-encode $2^n$ datum, a $n$-qubit discriminator
is fed by a generator with $2^n$ qumodes. In
Figure~\ref{fig:generatorqumodetcircuit}, $n$ is two.

The generator is initialized with arbitrary displacements and rotation
angles ($\alpha$ and $\phi$). A gradient descent optimizer is used
to minimize the cost represented by the term $-p_F$. The outcome
of the optimization of the generator is two column vectors of
displacements and rotation angles, $2^n$ rows each, actualizing the
generator circuit such that the probability that fake data is
recognized as true is high. 
\begin{defn}[Generator optimization problem]
Given the quantum input state $\psi$, 
probability amplitude encoding fake navigation data, 
the discriminator $\mathcal{D}(\omega)$, 
actualized with rotation angle matrix $\omega$,
training the generator $\mathcal{G}(\alpha,\phi)$ is the optimization problem 
that consists of finding the column vectors of the rotation angles  
$\alpha$ and $\phi$ ($2^n$ rows each) 
that gives the smallest difference $-p_F$.
\end{defn}

The learning process is automatic. The output of
the photonic quantum circuit is classical and in the continuous
domain. It is directly usable by the adversary to generate 
fake navigation data during a covert attack.
The circuit complexity is although in $\mathcal{O}(2^{n})$.

\section{Performance}
\label{sec:performance}

\begin{figure}[!b]
\centering
\includegraphics[width=1\columnwidth]{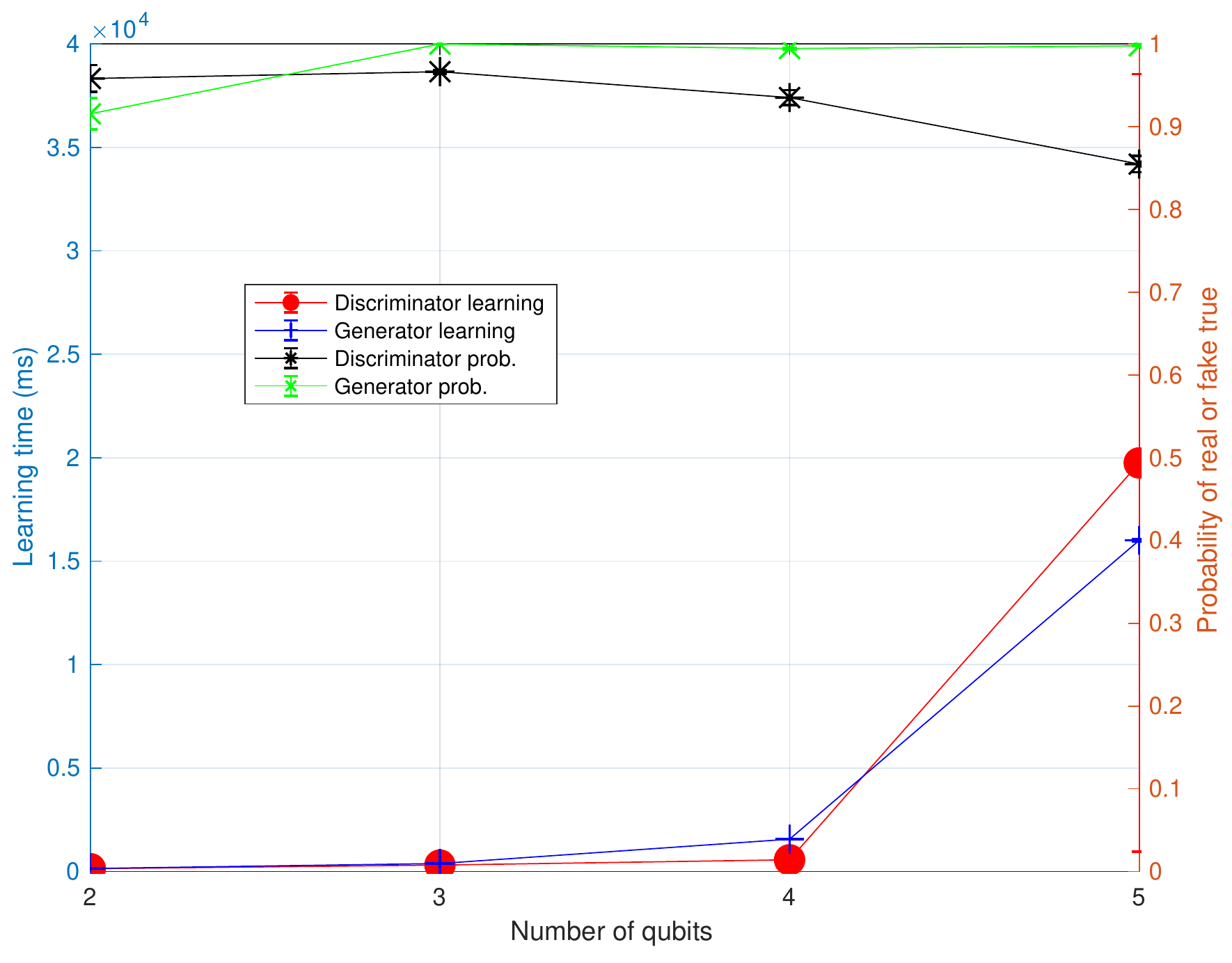}
\caption{Learning time (ms) versus the number of qubits for the
  discriminator and generator.\label{fig:learningtime}}
\end{figure}

The performance of the photonic-circuit design described in Section~\ref{sec:approach} has been validated through simulation on 
a classical computing platform. Simulations were conducted using 
an Intel Xeon 32-core 2.70~GHz server, with 256~GB of memory.
We generated genuine navigation for a \href{https://www.parrot.com/global/support/products/mambo-fpv}{Parrot Mambo} \gls*{mav}.
In the scenario, the \gls*{mav} takes off one~meter.
Does two circles on the horizontal plane, then lands.
The navigation data consists of $x$, $y$ and $z$ velocity triples.
The whole scenario generates less than $64$ real number values.

Figure~\ref{fig:learningtime} plots the discriminator and generator 
learning time (ms) versus the number of qubits available.
The $x$ axis represents the number of qubits.
The left $y$ axis refers to the learning time (ms).
The right $y$ axis shows the corresponding probability of real true,
for the discriminator, and probability of fake true, for the generator.
Hundred optimization iterations were done for each case.
Negligible error margins are included, but not visible since they 
are very tiny.
The discriminator is trained with six different genuine navigation data sets.
A navigation data set is picked at random at every optimization iteration.
The discriminator 
optimization time grows exponentially.
Due to the exponential complexity of the generator circuit (in $\mathcal{O}(2^{n})$),
the optimization time also grows exponentially. 
On our simulation platform, it becomes unpractical from six qubits.
The learning time becomes in the order of days.
Amplitude encoding has also $\mathcal{O}(2^{n})$ time complexity, 
but it is only executed once at the start of the optimization process.

\section{Conclusion}
\label{sec:conclusion}

We have investigated the use of \gls*{qgan} designs to generate fake \gls*{mav} navigation data. We assume the same approach to discriminate between genuine and fake \gls*{mav} navigation data. The goal pursued by the adversary is to generate fake data that is accepted
as true by a trained discriminator.
On the other hand, the discriminator must accept with high probabilities true navigation data and reject fake one. 
The elaborated quantum circuits have been evaluated running on a 
a classical computing platform.
As demonstrated in Figure~\ref{fig:learningtime},
the exponentially growing time complexity in the number of qubits is an obstacle to 
scalability.
We identified hurdles that must be overcome by the upcoming evolution of quantum machine learning.
The main hurdle for the adversary is the generation of navigation data in 
classical continuous domains, i.e., real numbers, and the cost of the 
transformation into the quantum format at every optimization iteration.
Further research is needed to improve and find alternatives to the design depicted in Figure~\ref{fig:generatorqumodetcircuit}.\\

\begin{footnotesize}
\noindent \textbf{Acknowledgments ---} Work partially supported by the Natural Sciences and Engineering Research Council of Canada (NSERC), the Cyber CNI Chair of the Institut Mines-T\'el\'ecom (cf. \url{https://www.chairecyber-cni.org/}), and the European Commission under grant agreement 830892 (H2020 SPARTA project, cf. \url{https://www.sparta.eu/}).
\end{footnotesize}

\bibliographystyle{plain}
\bibliography{main}

\end{document}